\documentclass[aps,pra,superscriptaddress,twocolumn]{revtex4-2}

\usepackage{graphicx}
\usepackage{amsmath}
\usepackage{amsthm}
\usepackage{amssymb}
\usepackage{hyperref}
\usepackage[utf8]{inputenc}
\usepackage{mathtools}
\usepackage[english]{babel}
\usepackage{bbm}
\hypersetup{colorlinks=true, linkcolor=blue, citecolor=blue, urlcolor=blue}
\usepackage{xcolor}
\usepackage[normalem]{ulem}
\usepackage{braket}

\newcommand{\T}{\mathcal{T}}
\renewcommand{\P}{\mathcal{P}}

\newcommand{\qhat}{\hat{\mathbf{q}}}
\newcommand{\khat}{\hat{\k}}
\newcommand{\dhat}{\hat{\mathbf{d}}}

\newcommand{\zhat}{\hat{\mathbf{z}}}
\newcommand{\xhat}{\hat{\mathbf{x}}}
\newcommand{\yhat}{\hat{\mathbf{y}}}

\newcommand{\q}{\mathbf{q}}

\newcommand{\R}{\mathcal{R}}
\renewcommand{\k}{\mathbf{{k}}}

\newcommand{\abs}[1]{\lvert #1 \rvert} 
\newcommand{\rcvect}[1]{
  \begin{pmatrix}#1\end{pmatrix}%
}

\DeclareMathOperator{\Tr}{Tr}
\DeclareMathOperator{\Imag}{Im}
\DeclareMathOperator{\Real}{Re}

\theoremstyle{definition}
\newtheorem{definition}{Definition}

\begin{document}

\title{Chirality-induced emergent spin-orbit coupling in topological atomic lattices}

\author{Jonah S. Peter}
\email{jonahpeter@g.harvard.edu}
\affiliation{Department of Physics, Harvard University, Cambridge, Massachusetts 02138, USA}
\affiliation{Biophysics Program, Harvard University, Boston, Massachusetts 02115, USA}
\author{Stefan Ostermann}
\affiliation{Department of Physics, Harvard University, Cambridge, Massachusetts 02138, USA}
\author{Susanne F. Yelin}
\affiliation{Department of Physics, Harvard University, Cambridge, Massachusetts 02138, USA}

\begin{abstract}
Spin-orbit coupling is of fundamental interest in both quantum optical and condensed matter systems alike. In this work, we show that optically induced electronic excitations in lattices of V-type atoms exhibit an emergent spin-orbit coupling when the geometry is chiral. This spin-orbit coupling arises naturally from the electric dipole interaction between the atomic sites and leads to a nontrivial topology for the lattice band structure. Using a general quantum optical model, we determine analytically the conditions that give rise to spin-orbit coupling and characterize the behavior under various symmetry transformations. We demonstrate that chirality-induced spin-orbit coupling can result from either the chirality of the underlying lattice geometry or the combination of an achiral lattice with a suitably chosen external quantization axis. We then discuss how these results are influenced by dissipation, which breaks time-reversal symmetry and illuminates the distinction between true and false chirality. Our results demonstrate that chiral atom arrays are a robust platform for realizing spin-orbit-coupled topological states of matter.

\end{abstract}

\maketitle


\section{Introduction}

The spin-orbit (SO) interaction refers to the coupling of a particle's intrinsic angular momentum to its motional degrees of freedom. For electrons in condensed matter systems, SO coupling can result in spin-momentum locking and nontrivial topological properties, resulting in new phases of matter including topological insulators~\cite{kane_quantum_2005, kane_z_2005}. Spin-orbit coupling is also relevant to the study of electron transport in chiral molecules, which have emerged as a promising new platform for spintronics devices~\cite{naaman_chiral_2019, michaeli_new_2017}. In particular, such phenomena are of considerable interest in the development of spin transistors, spin diodes, and other related technologies~\cite{zutic_spintronics_2004, manchon_new_2015, naaman_spintronics_2015, abendroth_spin_2019}. 

Angular momentum in photonic and quantum optical systems can also lead to phenomena analogous to conventional SO coupling. In these systems, the role of the electron spin is played by either the orthogonal polarizations of circularly polarized light or the internal angular momentum states of microresonators or atoms. Recently, various platforms have demonstrated the SO coupling of photons~\cite{lodahl_chiral_2017, bliokh_spinorbit_2015, bliokh_transverse_2015, bliokh_optical_2017}, as well as photonic analogues of the quantum spin hall effect~\cite{hafezi_robust_2011}. In cold atoms, time-reversal pairs of internal hyperfine states can act as pseudospin-1/2 systems that resemble electronic spin degrees of freedom~\cite{galitski_spinorbit_2013}. Coupling these states to coherent laser fields can produce synthetic SO potentials in ultracold Fermi gases~\cite{wang_spin-orbit_2012, cheuk_spin-injection_2012, liu_effect_2009} and Bose-Einstein condensates~\cite{lin_spinorbit-coupled_2011}. In addition to ultracold gases comprised of moving atoms, the hyperfine levels of atoms or atomlike emitters arranged in ordered lattices can also be mapped onto pseudospin-1/2 states~\cite{syzranov_spinorbital_2014}. Such atomic arrays support the transport of optically induced electronic excitations in a manner analogous to electrons in traditional crystal lattices~\cite{asenjo-garcia_exponential_2017, perczel_photonic_2017, perczel_topological_2017}, thus providing an orbital degree of freedom. The effective SO coupling produced in these systems can be leveraged as quantum simulation platforms to study SO dynamics in a highly tunable environment.

In this work, we demonstrate that the pseudospin-1/2 excitations of arrays of dipole-coupled atoms with a V-type level structure experience an effective SO coupling when the geometry is chiral. Our findings are distinct from those of previous works in that here the SO coupling results from the geometrical dependence of the dipole-dipole interaction and emerges naturally from the chirality of the system. We show that the associated photonic band structures of such arrays exhibit a finite spin texture and a topologically nontrivial Zak phase. Through a general symmetry analysis, we determine the analytical conditions that give rise to SO coupling in these arrays and explain how the SO coupling vanishes when the chirality is lost. We then discuss how these results are influenced by dissipation, which breaks time-reversal symmetry and illuminates the distinction between true and false chirality. Finally, we present a series of examples for different lattice configurations. Our results demonstrate that chiral atom arrays are a robust platform for realizing SO-coupled topological states of matter.

The remainder of the manuscript is organized as follows. Section~\ref{sect:formalism} describes the theoretical model used to encode the pseudospin-1/2 degree of freedom in atomic arrays. Section~\ref{sect:SOC} derives the necessary requirements for emergent SO coupling in terms of a  broken ``spin inversion" symmetry. Section~\ref{sect:symm} relates spin inversion symmetry to more conventional symmetry operations, including rotations, reflections, inversions, and time-reversal. Section~\ref{sect:lat} provides examples of chiral lattices, characterized by emergent SO coupling and topologically nontrivial band structures. Section~\ref{sect:setups} demonstrates how an achiral lattice, when combined with a suitable orientation of the atomic quantization axis, can also yield SO coupling and nontrivial topology. Section~\ref{sect:conclusion} summarizes the results and proposes suggestions for future work.

\section{Theoretical Formalism\label{sect:formalism}}

\begin{figure}
\centering 
\renewcommand\figurename{FIG.}
\includegraphics[width=\columnwidth]{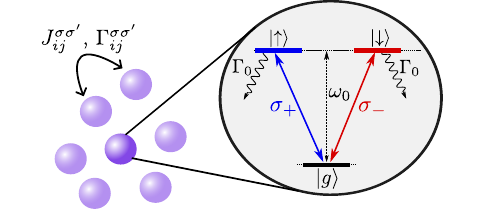}
\caption{Schematic of the V-type level structure considered in this work. The excited state orbitals have resonant frequency $\omega_0$ and spontaneous emission rate $\Gamma_0$. Emitters are coupled with coherent and dissipative hopping rates $J_{ij}^{\sigma\sigma'}$ and $\Gamma_{ij}^{\sigma\sigma'}$, respectively. 
}
\label{fig:level_structure}
\end{figure}

We consider ordered arrays of dipole-coupled quantum emitters (e.g., atoms, molecules, or quantum dots) in free space at fixed positions within the laboratory frame. We assume the emitters are tightly trapped at their respective locations such that the motional degrees of freedom can be neglected. The pseudospin degree of freedom for each emitter is encoded in a V-type level structure consisting of a single ground state $\ket{g}$ and two hyperfine states $\ket{\uparrow}$ and $\ket{\downarrow}$ corresponding to the two orthogonal polarizations of circularly polarized light (Fig. \ref{fig:level_structure}). The bare hyperfine states are assumed to have identical resonance frequencies $\omega_0 = 2 \pi c / \lambda_0$ where $\lambda_0$ is the wavelength of each optical transition and $c$ is speed of light in vacuum.

\subsection{Dipole-dipole interactions}

The transport of optically induced electronic excitations between quantum emitters in free space involves long-range interactions mediated by a radiation field. It is convenient to trace out the field degrees of freedom in the Born and Markov approximations to obtain an effective descriptive in terms of the matter operators only~\cite{lehmberg_radiation_1970, lehmberg_radiation_1970_2, asenjo-garcia_exponential_2017, breuer_theory_2010, scully_quantum_2008}. In free space, the effective interactions between quantum electric dipoles at points $\mathbf{r}_i$ and $\mathbf{r}_j$ are determined by the dyadic Green's tensor
\begin{multline}\label{eq:greens}
\mathbf{G}(\mathbf{r}_{ij}, \omega_0) = \frac{e^{ik_0 r_{ij}}}{4\pi k_0^2 r_{ij}^3}  \Big[ (k_0^2 r_{ij}^2 + ik_0 r_{ij} - 1) \openone\\
 - (k_0^2r_{ij}^2 + 3ik_0 r_{ij} - 3) \frac{\mathbf{r}_{ij} \otimes \mathbf{r}_{ij}}{r_{ij}^2} \Big],
\end{multline}
where $\mathbf{r}_{ij} = \mathbf{r}_i - \mathbf{r}_j$, $r_{ij} = \abs{\mathbf{r}_{ij}}$, and $k_0 = \omega_0/c$ (see also Appendix~\ref{app:dipole}). The coherent and dissipative parts of the dipole-dipole interaction are then given by
\begin{align}
J^{\sigma\sigma'}_{ij} &= - \frac{3}{2} \lambda_0 \Gamma_0 \pmb{\varepsilon}^{\dagger}_{i\sigma} \cdot \Real \mathbf{G}(\mathbf{r}_{ij}, \omega_0) \cdot \pmb{\varepsilon}_{j\sigma'}\label{eq:J}\\
\Gamma^{\sigma\sigma'}_{ij} &= 3 \lambda_0 \Gamma_0 \pmb{\varepsilon}^{\dagger}_{i\sigma} \cdot \Imag \mathbf{G}(\mathbf{r}_{ij}, \omega_0)\cdot \pmb{\varepsilon}_{j\sigma'},
\end{align}
where $\pmb{\varepsilon}_{i\sigma}$ is the circular polarization vector for orbital $\sigma \in \{\uparrow, \downarrow\}$ on emitter $i$, $\Gamma_0 \equiv \Gamma_{ii}^{\sigma\sigma} = \omega_0^3 \lvert \pmb{\wp}_{i\sigma} \rvert ^2 / (3 \pi \hbar \epsilon_0 c^3)$ is the spontaneous emission rate associated with each excited state orbital, $\pmb{\wp}_{i\sigma}$ is the transition dipole matrix element vector, and $\epsilon_0$ is the vacuum permittivity. The interactions therefore depend only on the scalar distance between the emitters and on the relative orientations of the polarization vectors. These polarization vectors are defined with respect to a quantization axis $\qhat$ about which the optically excited orbitals are circularly polarized. In practice, this quantization axis can be fixed in the laboratory frame by an external electric or magnetic field. The orientation of the quantization axis relative to the symmetry planes of the lattice is directly related to the emergence of SO coupling in chiral systems (Section~\ref{sect:SOC}). For general $\qhat$, the polarization vectors for left and right circularly polarized excitations are given by
\begin{equation}\label{eq:pols}
    \pmb{\varepsilon}_{\uparrow \downarrow} = \frac{1}{\sqrt{2}} (\hat{\mathbf{d}}_1 \pm i \hat{\mathbf{d}}_2),
\end{equation}
where $\uparrow$ ($\downarrow$) corresponds to $+$ ($-$), $\hat{\mathbf{d}}_1$ and $\hat{\mathbf{d}}_2$ denote the orthonormal vectors defining the polarization plane, and $\hat{\mathbf{d}}_1 \times \hat{\mathbf{d}}_2 = \qhat$. 

\subsection{Open system Hamiltonian\label{sect:open}}

Throughout this work, we focus on the single-excitation subspace which is sufficient to observe chirality-induced SO coupling. The unitary dynamics for an arbitrary arrangement of $N$ V-type quantum emitters interacting via the dipole-dipole interactions described above are given by the Hamiltonian
\begin{equation}\label{eq:general_ham}
    H = \sum_{i=1}^N \sum_{\sigma} \omega_0 \ket{\sigma_i}\bra{\sigma_i} + \sum_{i\neq j} \sum_{\sigma,\sigma'} J_{ij}^{\sigma \sigma'} \ket{\sigma_i}\bra{\sigma'_j}
\end{equation}
(we set $\hbar \equiv 1$ here and throughout). Here, the state $\ket{\sigma_i} = \ket{g,g,...,\sigma,...,g}$ denotes a single excitation in orbital $\ket{\sigma}$ at site $i$ with all the other emitters in the ground state. The non-unitary contributions of collective dissipation and single emitter spontaneous emission are included via the anti-Hermitian term
\begin{equation}\label{eq:H_A}
    H_A =  - \frac{i}{2} \sum_{i, j}^N \sum_{\sigma,\sigma'} \Gamma_{ij}^{\sigma \sigma'} \ket{\sigma_i}\bra{\sigma'_j}.
\end{equation}
The total non-Hermitian effective Hamiltonian describing the full open system dynamics is then
\begin{align}\label{eq:H_eff}
    H_\mathrm{eff} &= H + H_A \nonumber \\
    &= \sum_{i=1}^N \sum_{\sigma} \omega_0 \ket{\sigma_i}\bra{\sigma_i} -\frac{3}{2} \lambda_0 \Gamma_0 \sum_{i\neq j} \sum_{\sigma,\sigma'} G_{ij}^{\sigma \sigma'} \ket{\sigma_i}\bra{\sigma'_j},
\end{align}
where $G_{ij}^{\sigma \sigma'} = \pmb{\varepsilon}^{\dagger}_{i\sigma} \cdot \mathbf{G}(\mathbf{r}_{ij}, \omega_0) \cdot \pmb{\varepsilon}_{j\sigma'}$. We may further trace over the ground state of each emitter and denote the excited states using the basis vector mapping $\ket{\uparrow} = \rcvect{1 & 0}^T$, $\ket{\downarrow} = \rcvect{0 & 1}^T$ such that the circularly polarized excitations at each emitter site behave as pseudospin-1/2 degrees of freedom characterized by the $2 \times 2$ Pauli matrices. The operator $\sigma_z = \ket{\uparrow}\bra{\uparrow} - \ket{\downarrow}\bra{\downarrow}$ then quantifies the relative spin population in each emitter.

Finally, it is useful to define a set of collective operators that act on the spin indices of each emitter simultaneously. We denote
\begin{equation}
\mathcal{S}_\alpha \equiv \openone \otimes \sigma_\alpha
\end{equation}
for $\alpha \in \{x,y,z\}$, where $\openone$ is the identity matrix acting on the spatial indices. The pseudospin operator for a delocalized state extending across multiple emitters then follows simply as $\mathcal{S}_z$.

\subsection{Photonic band structures}

In order to characterize the SO-coupling properties of pseudospin-1/2 atomic lattices, we will assess the photonic band structures obtained by transforming the real-space Hamiltonian into momentum space. For simplicity, we limit the discussion to lattices that are periodic only in one direction. In this case, the site index for a non-Bravais lattice composed of $\mathcal{N}$ sublattices can be decomposed into $i=(m, \mu)$ where $m \in [1, M]$ indexes the unit cell along the axis of periodicity and $\mu \in [1, \mathcal{N}]$ denotes the sublattice index. In the limit of large $M$, the substitution $\ket{\sigma_i} = \ket{m, \mu, \sigma} = (1/\sqrt{M}) \sum_\k \exp{(i\k\cdot\mathbf{r}_{m\mu})} \ket{\k, \mu, \sigma}$ for quasimomentum $\k$ yields $H_{\mathrm{eff}} = \sum_{\k} H(\k)$, where the Bloch Hamiltonian
\begin{equation}\label{eq:k-space_H}
    H(\k) = \sum_{\mu,\nu} \sum_{\sigma,\sigma'} h_{\mu\sigma,\nu\sigma'}(\k) \ket{\k, \mu, \sigma} \bra{\k, \nu, \sigma'}
\end{equation}
has matrix elements \cite{perczel_photonic_2017, perczel_topological_2017}
\begin{equation}\label{eq:k-space_elements}
    h_{\mu\sigma,\nu\sigma'}(\k) = \omega_0 \delta_{\mu\nu} \delta_{\sigma\sigma'} +  \chi^I_{\mu\sigma,\nu\sigma'} + \chi^{II}_{\mu\sigma,\nu\sigma'}
\end{equation}
for 
\begin{align}
    \chi^I_{\mu\sigma,\nu\sigma'} &= -\frac{3}{2}\lambda_0 \Gamma_0 \sum_{\mathbf{R}_{\mu} \neq 0} e^{-i\k\cdot\mathbf{R}_{\mu}} G_{\sigma\sigma'}(\mathbf{R}_{\mu})\delta_{\mu\nu} \label{eq:chi_one}\\
    \chi^{II}_{\mu\sigma,\nu\sigma'} &=  -\frac{3}{2}\lambda_0 \Gamma_0 \sum_{\mathbf{R}_{\mu}} e^{-i\k\cdot \mathbf{R}_\mu} G_{\sigma\sigma'}(\mathbf{R}_\mu + \mathbf{n}_{\mu\nu})(1 - \delta_{\mu\nu}) \label{eq:chi_two}
\end{align}
(see also Appendix~\ref{app:band}). Here, the (infinite) set of $\mathbf{R}_{\mu}$ denotes vectors of the underlying Bravais lattice and $\mathbf{n}_{\mu\nu} = \mathbf{n}_{\mu} - \mathbf{n}_{\nu}$ is the basis vector pointing from sublattice $\nu$ to $\mu$ within a given unit cell. The off-diagonal terms $\chi^I$ and $\chi^{II}$ describe interactions between emitters on the same and different sublattices, respectively. The infinite sums are easily computed numerically and typically converge rapidly. Finally, the band structure is obtained by numerically diagonalizing $H(\k)$ at each point in the Brillouin zone. We note that, like the real-space Hamiltonian, $H(\k)$ is (in general) non-Hermitian.

\section{Conditions for Spin-Orbit Coupling\label{sect:SOC}}

The emergence of SO coupling within pseudospin-1/2 atomic lattices require a nontrivial spin texture for the Bloch modes. Put differently, it is a necessary and sufficient condition that the spin $\braket{\mathcal{S}_z}$ be nonzero at some point in the Brillouin zone in order to observe SO coupling. As a main result, we now determine analytically the conditions for $\braket{\mathcal{S}_z} \neq 0$. In particular, we will show that SO coupling emerges in systems that lack inversion symmetry about axes in the polarization plane, which motivates a generalized definition of chirality for pseudospin-1/2 atomic excitations. 

Quite generally, the spin of each Bloch mode is constrained to be zero if there exists a symmetry of the Bloch Hamiltonian that reverses the spin of each mode for all $\k$. We can define this ``spin inversion" symmetry through its action on the states associated with each unit cell. Each unit cell at position $\mathbf{R}$ along the Bravais lattice has $2\mathcal{N}$ states $\ket{\mathbf{R}, a}$. Here, $a=(\mu, \sigma)$ is a compound index denoting the combination of sublattice $\mu$ and spin $\sigma$. The spin inversion operator can then be written as
\begin{equation}\label{eq:W}
    W \ket{\mathbf{R}, a} = \sum_b W_{ab} \ket{\mathbf{R}, b},
\end{equation}
where $W = V \otimes \sigma_x$ is a $2\mathcal{N} \times 2\mathcal{N}$ matrix and $V$ is an $\mathcal{N} \times \mathcal{N}$ unitary matrix acting on the sublattice indices. Note that this symmetry leaves the position vector $\mathbf{R}$ unchanged (the quasimomentum $\k$ is also unchanged by virtue of the Fourier transform). In other words, $W$ is an operator that reverses the spin but leaves the lattice geometry---including the position along the Bravais lattice---invariant, up to a unitary transformation of the sublattice indices.

To see that this symmetry prohibits spinful Bloch bands, consider a general Bloch Hamiltonian $H(\k)$ with orthonormal eigenstates $\ket{u(\k)}$ and corresponding eigenvalues $\varepsilon(\k)$. Bloch's theorem guarantees orthonormality~\cite{asenjo-garcia_exponential_2017}, and we do not require $H(\k)$ to be Hermitian or time-reversal invariant (see also Section~\ref{sect:TR}). If $W$ is a symmetry of $H(\k)$, then $[H(\k),W] = 0$ such that
\begin{equation}
    W H(\k) \ket{u(\k)} = H(\k) W \ket{u(\k)} = \varepsilon(\k) W \ket{u(\k)}.
\end{equation}
The states $\ket{u(\k)}$ and $\ket{u'(\k)} \equiv W \ket{u(\k)}$ are therefore both eigenstates of $H(\k)$ with the same eigenvalue (for simplicity, we ignore the subtleties arising from multi-band gauge ambiguities that are resolved via the sewing matrix formalism~\cite{fu_topological_2007, bernevig_topological_2013}). Noting that $\sigma_x^\dag \sigma_z \sigma_x = - \sigma_z$, the spins of these Bloch states satisfy
\begin{align}\label{eq:no_spin}
    \braket{u'(\k) | \mathcal{S}_z | u'(\k)} &= \braket{u(\k) | W^\dag \mathcal{S}_z W | u(\k)} \nonumber \\
    &= - \braket{u(\k) | \mathcal{S}_z | u(\k)}.
\end{align}
Now, by orthonormality, we must have $\braket{u(\k) | u'(\k)} = \delta_{uu'}$. If $\braket{u(\k) | u'(\k)} = 1$, then there is no degeneracy and $\ket{u'(\k)} = e^{i\phi(\k)} \ket{u(\k)}$ for $\phi(\k) \in \mathbb{R}$. In this case, it follows trivially from Eq.~\eqref{eq:no_spin} that $\braket{\mathcal{S}_z} = 0$. If, on the other hand, $\braket{u(\k) | u'(\k)} = 0$, then the states are orthogonal with equal and opposite spin. However, because the states are degenerate, the linear combinations $\ket{u_\pm(\k)} \equiv (1/\sqrt{2}) (\ket{u(\k)} \pm \ket{u'(\k)})$ are also eigenstates of $H(\k)$ with the same eigenvalue. These superposition states satisfy $\braket{\mathcal{S}_z} = 0$ by construction. It follows that if $W$ is a symmetry of the Bloch Hamiltonian, then one may always construct a basis such that all Bloch modes have zero spin.

The remaining task is to relate this result back to the geometrical properties of the system. For $W$ to be a symmetry of the Bloch Hamiltonian, there must exist a unitary transformation that relates the original and spin-flipped configurations while preserving the mutual orientation of the basis and Bravais lattice vectors relative to the quantization axis $\q$. If such a transformation does not exist, then the spin inversion symmetry of the combined lattice-quantization axis system is broken. We may therefore take this as the \emph{definition} of chirality for pseudospin-1/2 systems. Indeed, Refs.~\cite{ayuso_ultrafast_2022, ordonez_generalized_2018} found that any chiral observable in the electric dipole approximation must be defined relative to an external vectorial quantity (in this case the quantization axis). In the following section, we will demonstrate how this generalized definition of chirality has a straightforward interpretation in terms of orthogonal transformations of the real space lattice geometry.

\section{Symmetry Analysis of Spin Bands\label{sect:symm}}

In this section, we demonstrate how spin inversion symmetry can be satisfied under a variety of different conditions depending on the geometrical and time-reversal properties of the underlying lattice. These concepts are then applied to specific lattice geometries in Sections~\ref{sect:lat} and \ref{sect:setups} in order to demonstrate emergent SO coupling (or lack thereof) through pedagogical examples.

\subsection{Orthogonal group symmetries\label{sect:ortho}}

In order to study how the condition for nontrivial spin textures relates back to the geometry of the system, we now consider transformations under representations of the orthogonal group $O(3)$ (see also Appendix~\ref{app:ortho} for more detail). If the unit cell is invariant under an orthogonal transformation $U$, then the operator corresponding to this symmetry may be written as~\cite{fu_topological_2007}
\begin{equation}
    U \ket{\mathbf{R}, a} = \sum_b U_{ab} \ket{U\mathbf{R}, b},
\end{equation}
where $U$ is a $2\mathcal{N} \times 2\mathcal{N}$ unitary matrix. The particular form of $U$ depends on the symmetry operation in question and on the structure of the unit cell. Nevertheless, a number of general relations can be deduced that apply to all Hamiltonians of the form~\eqref{eq:H_eff}.

We are particularly interested in transformations that satisfy Eq.~\eqref{eq:W} for spin inversion symmetry. The simplest case occurs for lattices that possess a mirror plane. In this case, the basis vectors $\mathbf{n}_{\mu}$ exist in mirror-symmetric pairs such that $\mathbf{n}_{\nu} = \mathcal{R}_M \mathbf{n}_{\mu}$ for reflection operator $\mathcal{R}_M$ through the plane (note also that $\mu = \nu$ when the lattice vectors $\mathbf{R}_\mu$ lie in the mirror plane). If this mirror plane also contains the quantization axis, then the corresponding operator (written as $\R_{\q,M}$) acts on the basis states as
\begin{equation}\label{eq:qm}
    \R_{\q,M} \ket{\mathbf{R}, \mu, \sigma} = \ket{\R_{\q,M} \mathbf{R}, \nu, \bar{\sigma}},
\end{equation}
where $\ket{\bar{\sigma}} = \sigma_x \ket{\sigma}$ denotes the spin-flipped state [see Eq.~\eqref{eq:reflect}]. This transformation satisfies Eq.~\eqref{eq:W}, provided $\R_{\q,M} \mathbf{R} = \mathbf{R}$. That is, if the lattice possesses a mirror plane that contains the quantization axis \emph{and} the Bravais lattice vectors, then the spin of each Bloch mode is guaranteed to be zero.

The transformation~\eqref{eq:qm} is not the only form of $W$ that imposes trivial spin textures. Rotoreflections are also possible, so long as the combined operation satisfies Eq.~\eqref{eq:W}. To flip the spin, the rotation should be by an angle $\pi$ about an axis lying in the polarization plane [see also Eq.~\eqref{eq:rotate_d}]. Denoting this rotation as $R_\perp (\pi)$, the transformation
\begin{equation}
    R_{\perp}(\pi) \R_{\q}  \ket{\mathbf{R}, \mu, \sigma}
    = \ket{R_{\perp}(\pi) \R_{\q} \mathbf{R}, \nu, \bar{\sigma}}
\end{equation}
also fulfills Eq.~\eqref{eq:W} but for a broader class of lattice geometries that satisfy $R_{\perp}(\pi) \R_{\q} \mathbf{R} = \mathbf{R}$ and $R_{\perp}(\pi) \R_{\q} \mathbf{n}_\mu = \mathbf{n}_\nu$. The physical interpretation of this result is that the Bloch modes have zero spin when the polarization plane contains a symmetry axis of improper rotation.

The considerations above justify the notion of spin inversion symmetry breaking as a form of generalized chirality for pseudospin-1/2 systems. Whereas true chirality is usually defined as a lack of \emph{any} axis of improper rotation, here we only require that such an axis not lie in the polarization plane. The latter definition naturally encompasses the former, but also includes additional configurations where the chirality stems from the mutual orientation of the lattice vectors and the quantization axis, rather than from the lattice geometry alone (Section~\ref{sect:setups}).

\subsection{Time-reversal symmetry\label{sect:TR}}

In addition to the geometry of the system, the behavior under time-reversal ($\T$) also influences the spin properties of the system. The dipole-dipole interaction present in Eq.~\eqref{eq:general_ham} leads to the hopping of pseudospin-1/2 excitations between sites $i$ and $j$ with a rate determined by the Green's tensor \eqref{eq:greens}. This process neglects electronic exchange interactions, which is a good approximation when the spacing between adjacent emitters is much larger than the spatial extent of the atomic wavefunctions. In this case, the circularly polarized excitations at each emitter site can be described using bosonic statistics and with a bosonic time-reversal operator. Generally, a Hamiltonian $H$ is $\T$-invariant if and only if there exists a unitary operator $U_{\T}$ such that $\T H \T^{-1} = H$ for $\T = \mathcal{K} U_{\T}$, where $\mathcal{K}$ is the anti-unitary complex conjugation operator $\mathcal{K}: i \to -i$~\cite{ryu_topological_2010}. The particular form of $U_\T$ depends on the choice of basis. For Hamiltonians of the form~\eqref{eq:general_ham} written in the pseudospin basis defined in Section~\ref{sect:open}, the time-reversal operator is given by
\begin{equation}\label{eq:TR}
    \T = \mathcal{K} \mathcal{S}_x
\end{equation}
and acts explicitly on the basis states as 
\begin{equation}
    \T \ket{\mathbf{R}, \mu, \sigma} = \T \ket{\mathbf{R}, \mu} \otimes \rcvect{1 \\ 0} = \ket{\mathbf{R}, \mu} \otimes \rcvect{0 \\ 1} = \ket{\mathbf{R}, \mu, \bar{\sigma}}.
\end{equation}
We note that this definition in terms of the pseudospin basis states is perfectly consistent with the textbook definition of complex conjugation acting on spherical harmonics~\cite{sakurai_modern_2011}. Clearly, $\T$ is bosonic and satisfies $\T^2 = 1$. This represents an important distinction from traditional electronic systems and significantly influences the spin textures of the resulting photonic bands.   

In traditional band theory, spin-1/2 electrons obeying fermionic statistics exist as degenerate Kramers' pairs when $\mathcal{T}$ symmetry is preserved. If inversion symmetry is also present, this constraint requires at least a two-fold degeneracy at every point in the Brillouin zone. If inversion symmetry is broken, $\mathcal{T}$ invariance still requires this degeneracy be preserved at all $\mathcal{T}$ invariant quasimomenta. In either case, the spin of each excitation may be interpreted as a vector on the Bloch sphere. 

In bosonic bands, no such degeneracy is required. For the pseudospins described by Eq.~\eqref{eq:pols}, the polarization of each excitation is instead given by a vector on the Poincare sphere \cite{poincare_theorie_1892, collett_field_2005} with left and right circular polarizations residing at the north and south poles, respectively. In contrast to the fermionic case (where an equal superposition of $\uparrow$ and $\downarrow$ spins results in an equal magnitude spin-1/2 excitation pointing in the orthogonal plane), vectors residing along the equator of the Poincare sphere do not carry angular momentum. That is, an equal superposition of left and right circularly polarized excitations yields a linearly polarized excitation of pseudospin-0 [see Eq.~\eqref{eq:pols}]. Whereas the $\mathcal{T}$ operator acting on a fermionic system corresponds to a complete inversion of the Bloch vector through the origin, the bosonic $\mathcal{T}$ results in a reflection through the Poincare sphere equatorial plane and leaves vectors residing in this plane unchanged. 

For the Hamiltonians considered here, the Hermitian part~\eqref{eq:general_ham} is always $\mathcal{T}$ invariant. In later sections, we will consider the influence of broken $\T$ symmetry as induced by the anti-Hermitian part of $H_\mathrm{eff}$ [Eq.~\eqref{eq:H_A}]. To this end, it is instructive to present a more detailed description of the behavior under $\mathcal{T}$. In particular, direct application of Eq.~\eqref{eq:TR} demonstrates that the pseudospin operator is time-odd, obeying $\mathcal{T} \mathcal{S}_z \mathcal{T}^{-1} = - \mathcal{S}_z$. If the Hamiltonian is time-reversal invariant, then $[H, \mathcal{T}] = 0$ and 
\begin{equation}\label{eq:time_reversal}
    \mathcal{T} H \mathcal{T}^{-1} = \mathcal{K} \mathcal{S}_x H (\mathcal{K} \mathcal{S}_x)^{-1} = \mathcal{S}_x H^* \mathcal{S}^{-1}_x = H.
\end{equation}
A general Bloch Hamiltonian $H(\k) = e^{-i\k \cdot \mathbf{r}} H e^{i\k \cdot \mathbf{r}}$ then satisfies
\begin{equation}\label{eq:T_bloch}
    \mathcal{T} H(\k) \mathcal{T}^{-1} = e^{i\k \cdot \mathbf{r}} \mathcal{S}_x  H^* \mathcal{S}^{-1}_x e^{-i\k \cdot \mathbf{r}} = H(-\k),
\end{equation}
where $H(-\k) = \mathcal{S}_x H^*(\k) \mathcal{S}^{-1}_x$. The Brillouin zone contains a set of time-reversal invariant momenta, $\k=\Gamma^i$, where $-\Gamma^i = \Gamma^i + \mathbf{G}_\mu$ and  $\mathbf{G}_\mu$ is a reciprocal lattice vector, $\mathbf{G}_\mu \cdot \mathbf{R}_\mu = 2\pi$ (to avoid ambiguity with other quantities denoted by $\Gamma$, we will always use a superscript to denote time-reversal invariant momenta)~\cite{fu_topological_2007}. At these points, the Bloch Hamiltonian satisfies $[H(\Gamma^i), \T] = 0$. To see how the bosonic $\T$ influences the degeneracy of the resulting Bloch bands, we again consider  $\ket{u(\k)}$ as a general eigenstate of $H(\k)$ with eigenvalue $\varepsilon(\k)$. As in the previous section, the spin associated with this state is given by $\braket{\mathcal{S}_z}$. Away from the $\Gamma^i$ points, the spin of this state is, in general, nonzero (Section~\ref{sect:SOC}). Applying Eq.~\eqref{eq:T_bloch}, the Bloch Hamiltonian for a time-reversal invariant system satisfies
\begin{equation}\label{eq:same_energy}
    \mathcal{T} H(\k) \ket{u(\k)} = H(-\k) \mathcal{T} \ket{u(\k)} = \varepsilon(\k) \mathcal{T} \ket{u(\k)}
\end{equation}
such that $\ket{u'(-\k)} \equiv \mathcal{T} \ket{u(\k)}$ is an eigenstate of $H(-\k)$ with the same energy, $\varepsilon'(-\k) = \varepsilon(\k)$. The states at $\k$ and $-\k$ necessarily have opposite spin:
\begin{align}\label{eq:opposite_spin}
    \braket{u'(-\k) | \mathcal{S}_z | u'(-\k)} &= \braket{u(\k) | \mathcal{T}^\dag \mathcal{S}_z \mathcal{T} | u(\k)} \nonumber \\
    &= \braket{u^*(\k) | \mathcal{S}_x \mathcal{S}_z \mathcal{S}_x^{-1} | u^*(\k)} \nonumber \\
    &= - \braket{u(\k) | \mathcal{S}_z | u(\k)},
\end{align} 
where the second step follows from $\mathcal{S}_x = \mathcal{S}_x^\dag = \mathcal{S}_x^{-1}$ and the third step from the Hermiticity of $\mathcal{S}_z$. However, because $\mathcal{T}^2 = 1$, $\ket{u(\k)}$ and $\ket{u'(\k)}$ need not represent orthogonal states. In the absence of accidental degeneracies, these states are instead equal (up to a phase) such that each band is symmetric about $\Gamma^i$ in energy and antisymmetric in spin. The requirement at $\k = \Gamma^i$ then follows simply as
\begin{equation}
    \braket{u(\Gamma^i) | \mathcal{S}_z | u(\Gamma^i)} = -\braket{u(\Gamma^i) | \mathcal{S}_z | u(\Gamma^i)} = 0
\end{equation}
for each band individually, and no two-fold spin degeneracy is required.

In position space, the eigenstates are equal superpositions of states at $\k$ and $-\k$. The antisymmetric spin textures required by $\T$ invariance therefore dictate that all position space eigenstates have zero spin. However, this does \emph{not} preclude the emergence of nontrivial spin textures, which requires only that the spin be nonzero at a \emph{single} point in the Brillouin zone. If this point has nonzero dispersion, then the antisymmetric nature of the spin bands---together with the symmetric nature of the energy bands---implies SO coupling for all $\T$ invariant systems with nonzero spin. If $\T$ symmetry is broken, this antisymmetric condition need not apply, and the position space eigenstates may also exhibit nonzero spin. 

Finally, it should be noted that in the absence of any Zeeman splitting term, the Hermitian Hamiltonian~\eqref{eq:general_ham} is always $\T$ invariant, regardless of the lattice geometry. However, the non-Hermitian Hamiltonian $H_\mathrm{eff}$ explicitly breaks $\T$ symmetry through the anti-Hermitian term~\eqref{eq:H_A}, which introduces dissipation to the otherwise unitary system. Under certain conditions, SO coupling may emerge through $H_\mathrm{eff}$ but not through $H$: that is, by breaking $\T$ symmetry. In analogy with Barron~\cite{barron_true_1986}, we denote systems exhibiting SO coupling in the presence of $\T$ symmetry as ``truly chiral," and those for which this SO coupling vanishes in the presence of $\T$ symmetry as ``falsely chiral."

\subsection{Inversion and anti-inversion symmetries}

If the lattice contains an inversion center at position $\mathbf{R} = 0$, then the parity ($\P$) operator may be written as~\cite{fu_topological_2007},
\begin{equation}\label{eq:parity}
    \P \ket{\mathbf{R},a} = \sum_b \P_{ab} \ket{-\mathbf{R},b},
\end{equation}
where $P = V \otimes \openone$ for unitary $V$. That is, $\P$ acts by inverting the spatial coordinates of the unit cell, up to a unitary transformation on the sublattice indices. We require that $\P$ leave the spin indices invariant because angular momentum is an axial vector. In this case, the Bloch bands are inversion symmetric about $\k = \Gamma^i$, and the Bloch Hamiltonian satisfies $\P H(\k) \P^{-1} = H(-\k)$. Similarly, when the Hamiltonian is time-reversal invariant, the Bloch bands are symmetric in energy and antisymmetric in spin. Suppose now that the Hamiltonian is invariant not under $\P$ or $\T$ but under the unitary operator
\begin{equation}\label{eq:anti-parity}
    \bar{\P} = \P \otimes \sigma_x.
\end{equation}
$\bar{\P}$ then acts as the parity operator but treats the spin degree of freedom as a polar vector that is odd under inversion (for a 1D lattice, we may also interpret this operation as a series of rotations). Consequently, $[H, \bar{\P}]=0$ implies $\bar{\P} H(\k) \bar{\P}^{-1} = H(-\k)$ such that
\begin{equation}\label{eq:Pbar_spin}
    \bar{\P} H(\k) \ket{u(\k)} = H(-\k) \bar{\P} \ket{u(\k)} = \varepsilon(\k) \bar{\P} \ket{u(\k)}.
\end{equation}
In analogy with Eqs.~\eqref{eq:same_energy} and \eqref{eq:opposite_spin}, the states $\ket{u(\k)}$ and $\bar{\P} \ket{u(\k)}$ have equal energy and opposite spin on opposite sides of the Brillouin zone. The operator $\bar{\P}$ therefore acts as a sort of ``anti-inversion" by enforcing a spin antisymmetry about $\k=\Gamma^i$. Importantly, this symmetry is a property of the lattice geometry and the orientation of the quantization axis, and is independent of the behavior under $\T$.

\begin{figure}
\centering 
\renewcommand\figurename{FIG.}
\includegraphics[width=\columnwidth]{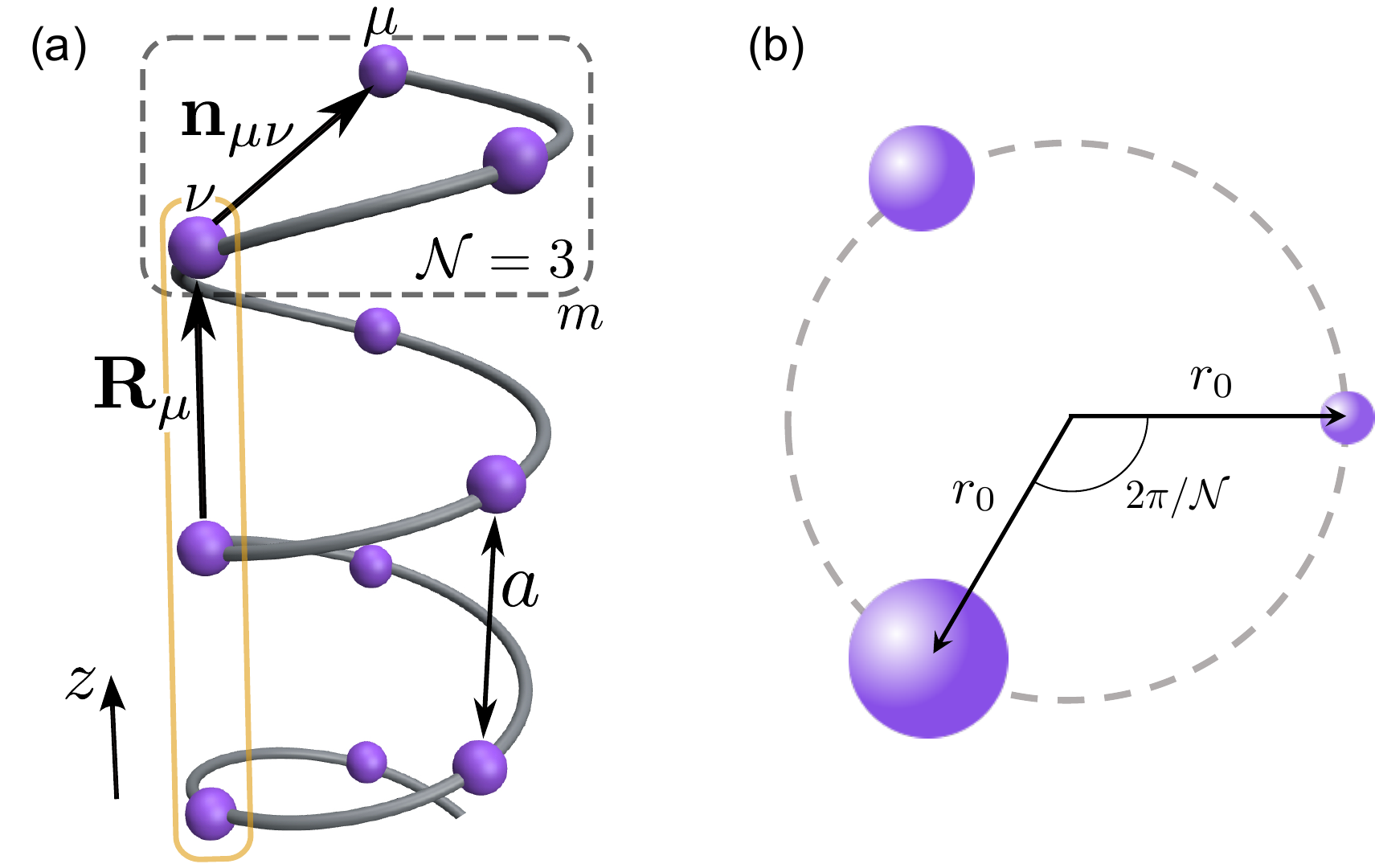}
\caption{(a) Schematic of the helical lattice geometry for pitch $a$ and $\mathcal{N} = 3$. The dashed grey rectangle denotes one unit cell, whereas the solid yellow rectangle denotes one sublattice. (b) Top-down view of the helical lattice. The sublattices are arranged with radial position $r_0$ and azimuthal separation $2 \pi / \mathcal{N}$.}
\label{fig:helix_unit_cell}
\end{figure}

\section{Chiral lattices\label{sect:lat}}

We are now in a position to study specific lattice geometries. We first consider ``chiral lattices," which we define as those exhibiting either true or false chirality for all orientations of $\qhat$. The quintessential chiral structure is a right circular helix, which serves as a paradigmatic example. This structure is periodic along its longitudinal helical axis which, without loss of generality, we choose to be parallel to the $z$-axis (Fig.~\ref{fig:helix_unit_cell}). The lattice vectors of the underlying Bravais lattice are then given by $\mathbf{R}_\mu = al \hat{\mathbf{z}}$ for lattice constant $a$ and $l \in \mathbb{Z}$. For a right-handed helix periodic along the $z$-axis with radius $r_0$, pitch $a$, and $\mathcal{N}$ emitters per unit cell, the emitter positions are given by
\begin{equation}\label{eq:pos}
\mathbf{r}_{j} = r_0 \cos(\phi_j) \hat{\mathbf{x}} + r_0 \sin(\phi_j) \hat{\mathbf{y}} + \frac{a \phi_j}{2\pi} \hat{\mathbf{z}},
\end{equation}
where $\phi_j = 2 \pi (j-1) / \mathcal{N}$. The relative coordinate between emitters $i$ and $j$ can then be written as $\mathbf{r}_{ij} = x_{ij} \hat{\mathbf{x}} + y_{ij} \hat{\mathbf{y}} + z_{ij} \hat{\mathbf{z}}$ for
\begin{align}\label{eq:relative}
   x_{ij} &= -2r_0 \sin\left[(\mu-\nu)\frac{\pi}{\mathcal{N}}\right] \sin\left[(\mu+\nu)\frac{\pi}{\mathcal{N}}\right] \nonumber\\
    y_{ij} &= 2r_0 \sin\left[(\mu-\nu)\frac{\pi}{\mathcal{N}}\right] \cos\left[(\mu+\nu)\frac{\pi}{\mathcal{N}}\right] \nonumber\\
    z_{ij} &= \left[(\mu-\nu)\frac{a}{\mathcal{N}} + al\right].
\end{align}
Here, $\mu \ (\nu) = i \ (j) \ \mathrm{mod} \ \mathcal{N}+1$ for integers $i,j \in [1, N]$ and $l$ denotes the number of unit cells between $i$ and $j$ (i.e., $l=0$ if $i$ and $j$ are in the same unit cell). We note that with this choice of coordinates, the $\mu=1$ sublattice lies on the positive $x$-axis. 

\begin{figure}
\centering 
\renewcommand\figurename{FIG.}
\includegraphics[width=\columnwidth]{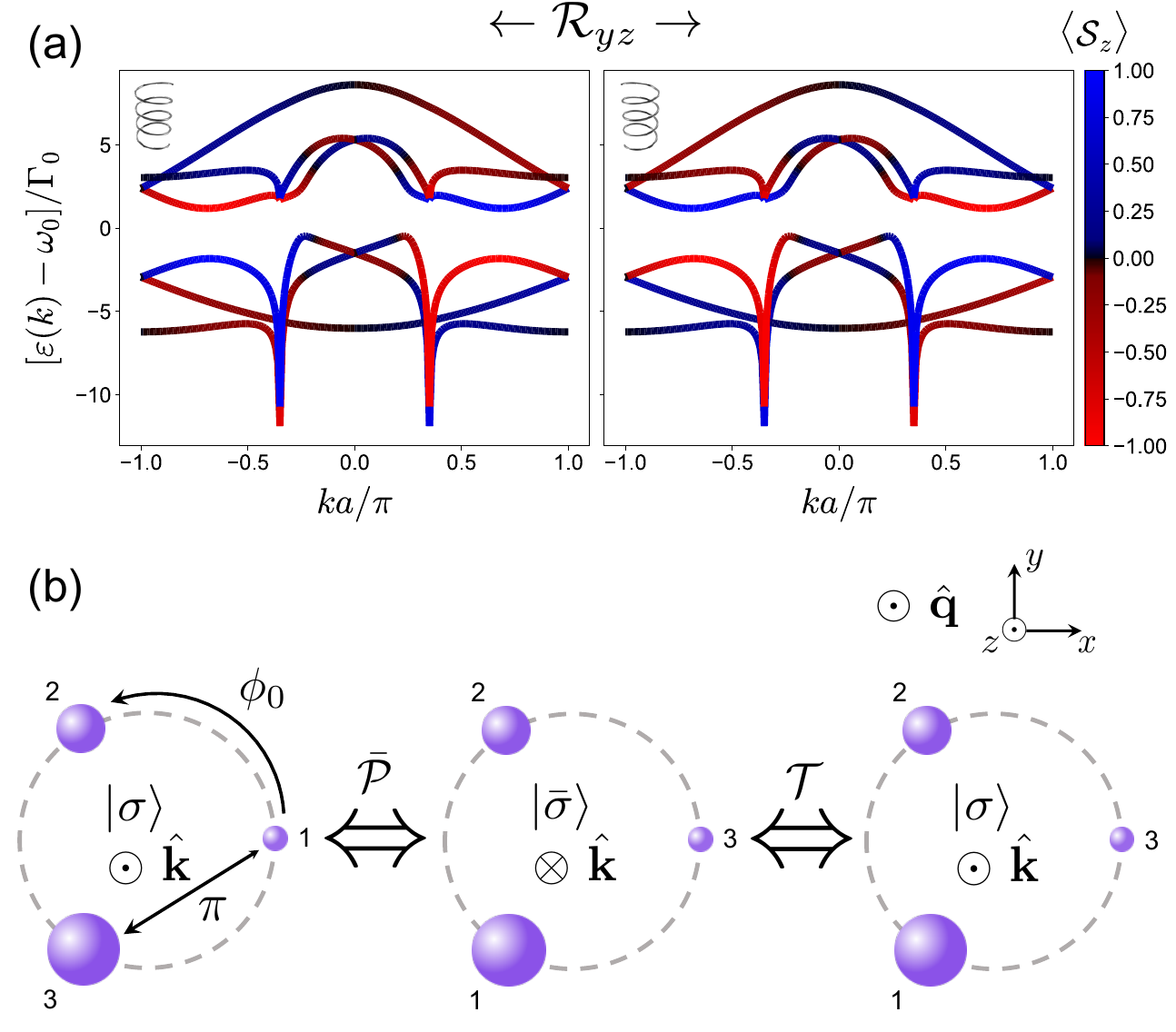}
\caption{(a) Band structures for the left-handed (left) and right-handed (right) helices when $\qhat = \zhat$. The spin bands are nontrivially antisymmetric about $\k=\Gamma^i$, irrespective of $\T$-invariance. The system exhibits true chirality with a quantized nontrivial Zak phase. Additional parameters: $\mathcal{N} = 3$, $r_0 = 0.05 \lambda_0$, $a = 0.175 \lambda_0$. (b) Top-down view of the helical unit cell for $\qhat = \zhat$. The $\mu=2$ sublattice serves as an anti-inversion for the lattice, and the corresponding transformation $\bar{\P}: (\khat, \sigma) \to (-\khat, \bar{\sigma})$ enforces antisymmetric Bloch bands. If dissipation is neglected, the spin antisymmetry is also protected by $\T$ symmetry.}
\label{fig:z_helix}
\end{figure}

\subsection{Longitudinal circular polarization\label{sect:long}}

As a demonstration of true chirality, we first orient the quantization axis to lie along the axis of periodicity ($\qhat = \zhat$). In this case, $r_0 = \varrho_i = \varrho_j$ and the spin-flip interaction is found to be
\begin{multline}
        G_{ij}^{\uparrow \downarrow} = \frac{3r_0^2 \lambda_0 \Gamma_0}{4\pi k_0^2 r_{ij}^5} \sin^2 \left (\frac{\phi_i - \phi_j}{2} \right ) e^{-i(\phi_i + \phi_j)} \\
    \times e^{ik_0 r_{ij}}\left (3 - k_0^2 r_{ij}^2 - i3k_0 r_{ij} \right ),
\end{multline}
where $\phi_i = \tan^{-1}(y_i/x_i)$. A reflection through the $y$-$z$ plane then takes $\phi_i \to - \phi_i$ such that $G_{ij}^{\uparrow \downarrow} \to G_{ij}^{\downarrow \uparrow}$ (see also Appendix~\ref{app:ortho}). The spin dynamics are therefore interchanged by reflections perpendicular to the polarization plane that transform the right-handed helix to its left-handed mirror image. 

The photonic band structures for left- and right-handed helices are shown in Fig.~\ref{fig:z_helix}(a). Because the helix geometry lacks an axis of improper rotation, the symmetry $[H(\k), W] \neq 0$ is broken for all unitary $W$ and the Bloch bands exhibit nontrivial spin textures. As a consequence of this broken symmetry, the longitudinally polarized helix supports bulk helical modes that experience SO coupling. This effect is illustrated by the spin antisymmetry of the band structures. Away from the points $\Gamma^i = 0, \pm \pi/a$, each Bloch mode acquires a finite group velocity $v(k) = \partial \varepsilon(k) / \partial k$ along the axis of periodicity. Because of the spin antisymmetry, these modes exhibit a helicity $\eta = \braket{\mathcal{S}_z} v / \abs{\braket{\mathcal{S}_z} v}$ that is symmetric about $\Gamma^i$. When dissipative interactions are neglected [i.e., the Hamiltonian is of the form \eqref{eq:general_ham}], this antisymmetry is protected by time-reversal symmetry. Invariance of the group velocity under reflection through the $y$-$z$ plane ($\mathcal{R}_{yz}$) then dictates that the two chiralities support equal and opposite helical modes. 

Specifically for $\qhat = \zhat$, the broken spin inversion symmetry persists regardless of whether the Hamiltonian is $\T$ invariant. When dissipative interactions are included, the Hermitian Hamiltonian~\eqref{eq:general_ham} is replaced by the non-Hermitian effective Hamiltonian~\eqref{eq:H_eff}. In this case, $\T$ symmetry is broken and Eq.~\eqref{eq:opposite_spin} no longer holds. However, the spin antisymmetry of the Bloch bands persists because the non-Hermitian Hamiltonian remains invariant under $\bar{\P}$. In other words, the $z$-polarized helix contains a 1D anti-inversion center at the center of the unit cell. 

To verify that the Hamiltonian is invariant under this transformation, we construct explicitly the associated matrix representation. The anti-inversion center lies at the azimuthal angle $\phi_0 = \pi (\mathcal{N}-1) / \mathcal{N}$ (as measured from the $x$-axis) along an axis that bisects the angle between sublattices $\mu = 1$ and $\mu = \mathcal{N}$. With this point chosen as the origin, the unit cell is symmetric under the combined operation of 1D spatial inversion and spin-flip. As demonstrated in Fig.~\ref{fig:z_helix}(b), this operation is equivalent to a $\pi$ rotation about the $\phi_0$ axis. $\bar{\P}$ therefore reverses the direction of $\mathbf{R}$ (or equivalently, $\k$) and flips the spin, while simultaneously exchanging the positions of the three sublattices (for odd $\mathcal{N}$, the central sublattice remains fixed). As such, the symmetry operation can be represented as $\bar{\P} = X \otimes \sigma_x$, where $X$ is the $\mathcal{N} \times \mathcal{N}$ antidiagonal exchange matrix acting on the sublattice indices. With the above choice of origin, the basis states of each unit cell then transform as
\begin{equation}\label{eq:anti}
    \bar{\P} \ket{\mathbf{R}, \mu, \sigma} = X \otimes \sigma_x \ket{-\mathbf{R}, \mu, \sigma}.
\end{equation}
One may verify explicitly that $[H_\mathrm{eff}, \bar{\P}] = 0$ and that the spin antisymmetry follows accordingly as in Eq.~\eqref{eq:Pbar_spin}. The net result is that the longitudinally polarized helix exhibits antisymmetric spin textures that are protected by anti-inversion symmetry (or equivalently, rotational symmetry). This property manifests even in the presence of time-reversal invariance: the system is truly chiral.

To examine the topological properties of the Bloch bands, we compute the Zak phase,
\begin{equation}
    \varphi = \oint_{\mathcal{C}} \Tr{[\mathbf{A}(\k)]} \cdot d\k.
\end{equation}
Here, $\mathbf{A}_{mn}(\k) = i \braket{u_{m\k} | \nabla_\k u_{n\k}}$ is the non-Abelian Berry connection matrix evaluated over a closed loop $\mathcal{C}$ around the first Brillouin zone (see Appendix~\ref{app:topo}). The Zak phase is defined modulo $2\pi$ and is quantized to either $0$ (trivial) or $\pm \pi$ (topologically nontrivial) when the Hamiltonian commutes with $\P$ or $\bar{\P}$. The finite spin textures exhibited by the longitudinally polarized helix induce a nontrivial topology in the energy bands. The associated SO coupled dynamics are therefore topologically protected by the chirality of the geometry. Because of the anti-inversion center at $\phi_0$, the Zak phase is quantized to $\pm \pi$.

\begin{figure}
\centering 
\renewcommand\figurename{FIG.}
\includegraphics[width=\columnwidth]{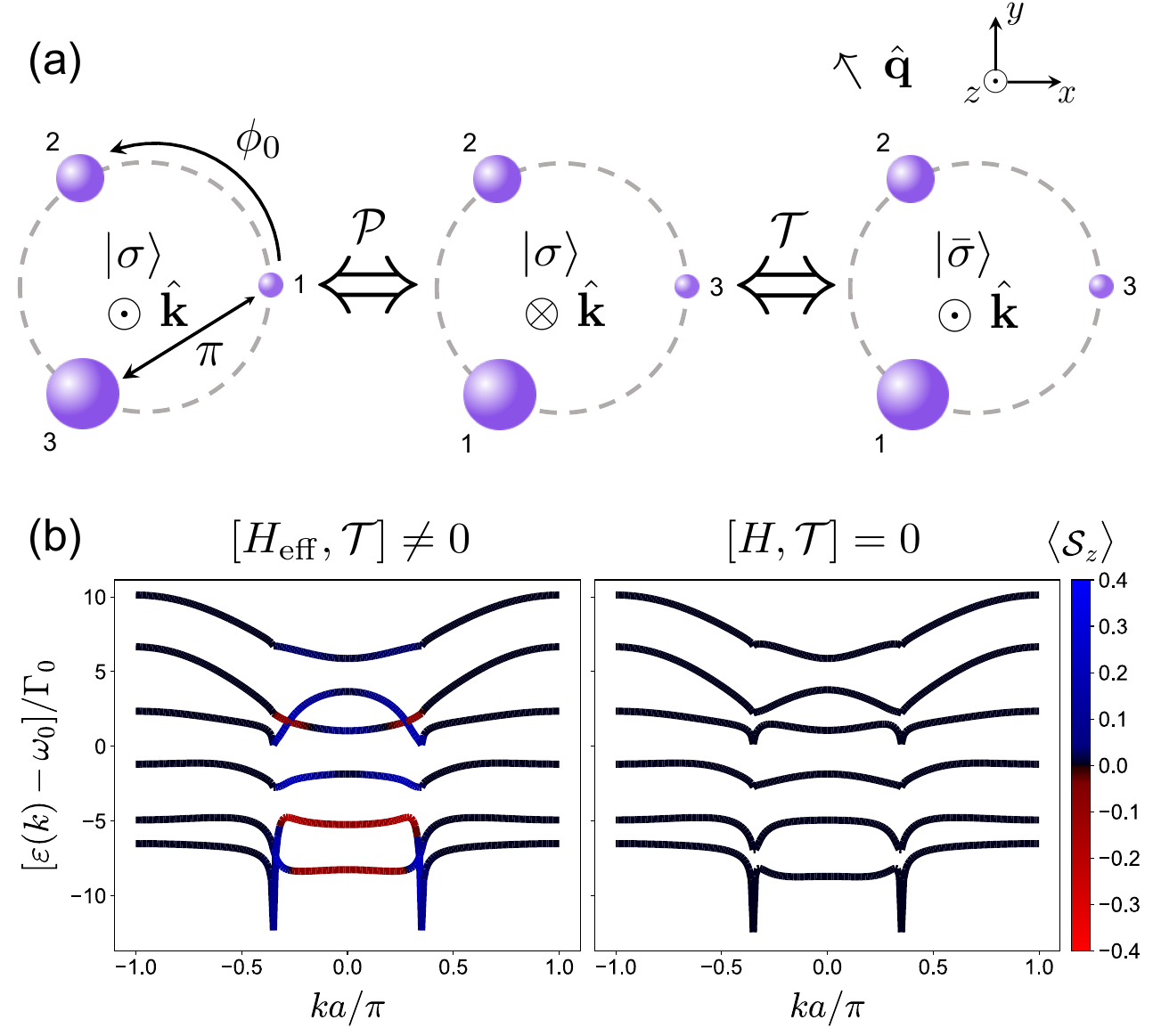}
\caption{(a) Top-down view of the helical unit cell for $\q$ oriented along the $\phi_0$ axis. The $\mu=1$ and $\mu=3$ sublattices are separated by an angle $\pi$ in the polarization plane such that the $\mu=2$ sublattice serves as a true inversion center. Inversion symmetry transforms $\P: (\khat, \sigma) \to (-\khat, \sigma)$, enforcing spin symmetric Bloch bands. If the Hamiltonian is also $\T$-invariant, the additional relation $\T: (\khat, \sigma) \to (-\khat, \bar{\sigma})$ requires $\braket{\mathcal{S}_z} = 0$. (b) Band structures for the $\phi_0$-polarized helix. The system is $\P$-invariant with a quantized Zak phase. When $\T$ symmetry is broken (left), the spin bands are symmetric about $\k=\Gamma^i$ and the Zak phase is topologically nontrivial. If $\T$ symmetry is restored (right), the combined $\P\T$ symmetry causes both the spin textures and the Zak phase to vanish. The system is therefore falsely chiral. Helix parameters are the same as in Fig.~\ref{fig:z_helix}.}
\label{fig:alpha_helix}
\end{figure}

\subsection{Transverse circular polarization}

The symmetry properties of the non-Hermitian Hamiltonian are altered when the quantization axis is oriented perpendicular to the axis of periodicity. As an example, we consider the helix of Eq.~\eqref{eq:pos} but with $\qhat$ pointing along the $\phi_0$ axis [Fig.~\ref{fig:alpha_helix}(a)]. Because the lattice geometry is unchanged, it remains true that there is no unitary $W$ enforcing $\braket{\mathcal{S}_z}=0$. However, the two-fold rotational symmetry of the lattice no longer corresponds to $\bar{\P}$. Instead, the $\pi$ rotation that swaps the $\mu=1$ and $\mu=3$ sublattices leaves the spin projection along $\qhat$ invariant. Consequently, the system exhibits a true inversion center and is parity symmetric with
\begin{equation}\label{eq:symm}
    \P \ket{\mathbf{R}, \mu, \sigma} = X \otimes \openone \ket{-\mathbf{R}, \mu, \sigma}.
\end{equation}
The spin bands are therefore symmetric about $\k=\Gamma^i$ [Fig.~\ref{fig:alpha_helix}(b)], and the Zak phase is quantized to $\pm \pi$. 

If dissipative interactions are neglected, then the Hamiltonian also commutes with $\T$, and the spin bands must also be antisymmetric [Eq.~\eqref{eq:opposite_spin}]. The combined $\P\T$ symmetry then forces $\braket{\mathcal{S}_z}=0$ for all $\k$, and the SO coupling is lost. In turn, the system becomes topologically trivial and the Zak phase is zero for each band. With this choice of quantization axis, the helix is only falsely chiral.

\begin{figure}
\centering 
\renewcommand\figurename{FIG.}
\includegraphics[width=\columnwidth]{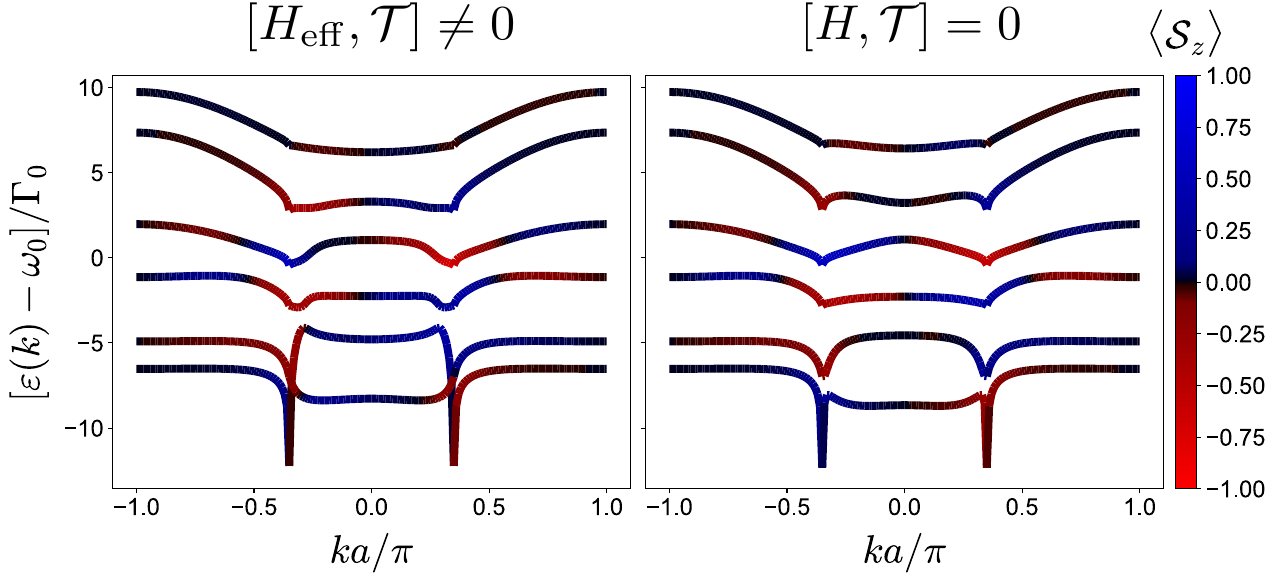}
\caption{Band structures for the transversely polarized helix in the absence of both $\P$ and $\bar{\P}$ invariance. When $\T$ symmetry is broken (left) the spin bands are neither symmetric nor antisymmetric about $\k=\Gamma^i$. When $\T$ symmetry is restored (right) the spin bands become antisymmetric. In both cases, the Zak phase is not quantized. Helix parameters are the same as in Figs.~\ref{fig:z_helix} and \ref{fig:alpha_helix}.}
\label{fig:45_helix_bands}
\end{figure}

For completeness, we also present the case where $\qhat$ does not correspond to a symmetry axis of the unit cell (e.g., lies at an angle $\pi/4$ from the $x$-axis in the $x$-$y$ plane). This configuration is not invariant under $\P$ or $\bar{\P}$. Hence, when $\T$ symmetry is broken, the spin bands are neither symmetric nor antisymmetric (Fig.~\ref{fig:45_helix_bands}). If however dissipation is neglected, then the Hamiltonian is $\T$-invariant and the antisymmetric spin textures are restored. In either case, the Zak phase is not quantized.

\section{Chiral setups\label{sect:setups}}

\begin{figure*}
\centering 
\renewcommand\figurename{FIG.}
\includegraphics[width=\textwidth]{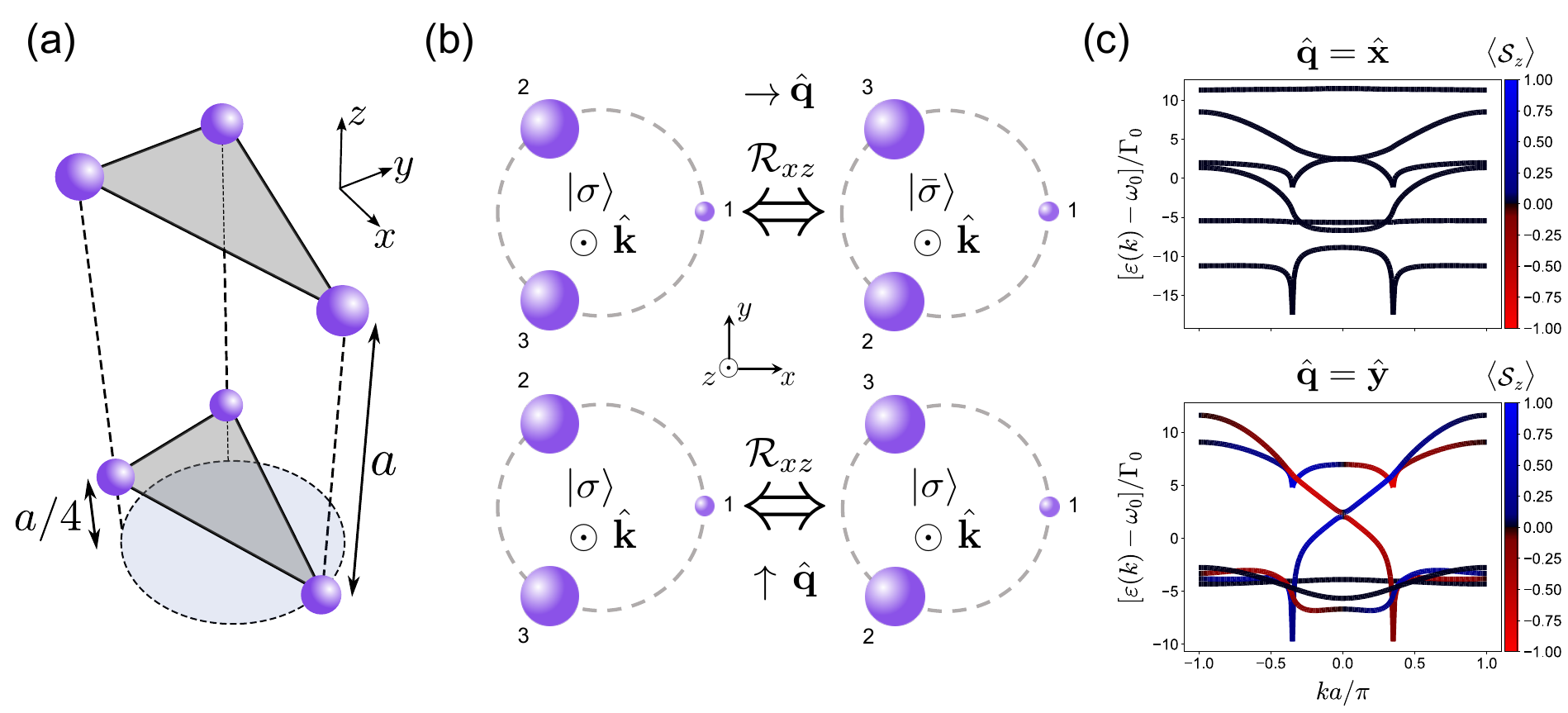}
\caption{(a) Lattice geometry for an oblique triangular prism. Each triangular face represents a single unit cell. The faces are assumed to be isosceles with the $x$-$z$ plane chosen as the mirror plane. (b) Top-down view of the oblique triangular prism unit cell for $\qhat = \xhat$ (top) and $\qhat = \yhat$ (bottom). (c) Band structures for the oblique triangular prism. When $\q$ is oriented in the mirror plane (left), the spin textures vanish irrespective of the behavior under $\T$. If $\q$ is instead oriented orthogonal to the mirror plane, then the SO coupling can be nonzero. In both cases, the Zak phase is not quantized. Additional geometrical parameters are the same as in Figs.~\ref{fig:z_helix}, \ref{fig:alpha_helix}, and \ref{fig:45_helix_bands}.}
\label{fig:triangle}
\end{figure*}

The spin inversion symmetry breaking required for finite SO coupling need not come from the lattice geometry alone.
If the lattice contains a mirror plane, there may still exist a choice of quantization axis such that the Bloch bands exhibit nontrivial spin textures. We refer to this scenario as a ``chiral setup" because not all orientations of $\qhat$ satisfy $[H(\k), W] \neq 0$ for unitary $W$. 

A concrete example is demonstrated by a lattice arranged into an oblique triangular prism. If the triangular faces are isosceles, then the lattice possesses a mirror plane along the axis of periodicity [Fig.~\ref{fig:triangle}(a)]. For simplicity, we choose the emitter positions to be circumscribed about a right circular cylinder such that $x_{ij}$ and $y_{ij}$ are given by Eq.~\eqref{eq:relative}, and
\begin{equation}
z_{ij} = 
\begin{cases} 
      0, & \mu,\nu \in \{2,3\} \\
      a/2, & \mu > \nu \\
      -a/2, & \mu < \nu
   \end{cases}
\end{equation}
where $\mu$ and $\nu$ are defined as before. For $\qhat = \xhat$, the quantization axis lies in the mirror plane [Fig.~\ref{fig:triangle}(b)], and the reflection operator $\R_{xz}$ acts as in Eq.~\eqref{eq:qm}. This mirror reflection commutes with both the Hermitian and non-Hermitian Hamiltonians such that the geometry is spin inversion symmetric irrespective of the behavior under $\T$. It follows that $\braket{\mathcal{S}_z}$ must vanish for all $\k$, and the system is not chiral at all.

By contrast, if the quantization axis is instead oriented along the $y$-axis, then $\R_{xz}$ leaves the spin invariant, and the SO coupling is therefore preserved. Because the unit cell does not contain a center of (anti-)inversion, the Bloch bands are not fully (anti)symmetric and the Zak phase is not quantized. Nevertheless, the antisymmetry of the nontrivial spin textures is consistent with $\T$ symmetry, and thus the system is truly chiral.

\section{Conclusions\label{sect:conclusion}}

In summary, we have demonstrated that the pseudospin-1/2 excitations admitted by chiral arrangements of V-type quantum emitters experience and emergent SO coupling and nontrivial topology. We have defined the conditions for SO coupling in terms of a broken spin inversion symmetry, and provided a complete characterization under different symmetry transformations. Our results describe a general phenomenon that is unique to chiral systems and can be achieved either with a chiral lattice, or via suitable orientation of the quantization axis. Because the emergence of SO coupling is dictated by the symmetries of the Bloch bands, the symmetry analysis given above is complete within the single excitation regime. Nevertheless, an interesting follow-up would be to analyze the dynamics of such arrays in the multi-excitation regime. In this case, the V-type level structure imposes a hard-core boson constraint in which each atomic orbital is, at most, singly occupied. The ensuing non-linearity induced by the hard-core interaction is expected to result in modifications of the transport phenomena. In this regard, chiral atom arrays could serve as a promising platform for new quantum optical devices and for studying atomic analogues of many-body topological physics in condensed matter systems.   


\section{Acknowledgements}
The authors are grateful to Mikhail D. Lukin and Jonathan Simon for suggestions on potential experimental implementations. S.O. is supported by a postdoctoral fellowship of the Max Planck-Harvard Research Center for Quantum Optics. All authors acknowledge funding from the National Science Foundation (NSF) via the Center for Ultracold Atoms (CUA) Physics Frontiers Centers (PFC) program and via PHY-2207972, as well as from the Air Force Office of Scientific Research (AFOSR).

\onecolumngrid
\appendix

\section{The electromagnetic Green's tensor\label{app:dipole}}

In free space, the effective interactions between quantum emitters are determined by the dyadic Green's tensor, $\mathbf{G}(\mathbf{r}, \mathbf{r'}, \omega)$, which is the solution to the wave equation
\begin{equation}
    \nabla^2 \mathbf{G}(\mathbf{r}, \mathbf{r'}, \omega) - \frac{\omega^2}{c^2} \mathbf{G}(\mathbf{r}, \mathbf{r'}, k) = \delta(\mathbf{r} - \mathbf{r'}) \openone
\end{equation}
for observational coordinates $\mathbf{r}$, source coordinates $\mathbf{r}'$, and frequency $\omega$. For the case where the emitters located at positions $\mathbf{r}_j$ are well approximated by point electric dipoles, the Green's tensor between emitters $i$ and $j$ depends only on the relative coordinate $\mathbf{r}_{ij} = \mathbf{r}_i - \mathbf{r}_j$. In the Born and Markov approximations, the Green's tensor may further be regarded as dispersionless and is given by~\cite{lehmberg_radiation_1970, lehmberg_radiation_1970_2, asenjo-garcia_exponential_2017, breuer_theory_2010, scully_quantum_2008}
\begin{equation}
\mathbf{G}(\mathbf{r}_{ij}, \omega_0) = \frac{e^{ik_0 r_{ij}}}{4\pi k_0^2 r_{ij}^3}  \Big[ (k_0^2 r_{ij}^2 + ik_0 r_{ij} - 1) \openone - (k_0^2r_{ij}^2 + 3ik_0 r_{ij} - 3) \frac{\mathbf{r}_{ij} \otimes \mathbf{r}_{ij}}{r_{ij}^2} \Big],
\end{equation}
with $r_{ij} = \abs{\mathbf{r}_{ij}}$ and $k_0 = \omega_0/c$. Consequently, the polarization (or photon ``spin") dependent dipole-dipole interaction between emitters $i$ and $j$ is given by $J^{\sigma\sigma'}_{ij} - (i/2) \Gamma^{\sigma\sigma'}_{ij}$, where~\cite{asenjo-garcia_exponential_2017}
\begin{align}
J^{\sigma\sigma'}_{ij} &= - \mu_0 \omega_0^2 \pmb{\wp}^{\dagger}_{i\sigma} \cdot \Real \mathbf{G}(\mathbf{r}_{ij}, \omega_0) \cdot \pmb{\wp}_{j\sigma'} \\
\Gamma^{\sigma\sigma'}_{ij} &= 2\mu_0 \omega_0^2 \pmb{\wp}^{\dagger}_{i\sigma} \cdot \Imag \mathbf{G}(\mathbf{r}_{ij}, \omega_0)\cdot \pmb{\wp}_{j\sigma'}
\end{align}
describe the coherent and dissipative parts of the interaction, respectively, and $\mu_0$ is the vacuum permeability.


\section{Photonic band structures for quasi-1D non-Bravais lattices\label{app:band}}

Beginning with the real-space non-Hermitian effective Hamiltonian, we expand the site index $i = (m,\mu)$ to yield
\begin{equation}
    H_{\mathrm{eff}} = \sum_{m=1}^{M} \sum_{\mu=1}^\mathcal{N} \sum_{\sigma} \omega_0 \ket{m, \mu, \sigma} \bra{m, \mu, \sigma} - \frac{3}{2}\lambda_0 \Gamma_0 \sum_{m,n=1}^{M} \sum_{\mu,\nu=1}^\mathcal{N} \sum_{\sigma,\sigma'} (1 - \delta_{mn} \delta_{\mu\nu}) G_{m\mu,n\nu}^{\sigma \sigma'} \ket{m, \mu, \sigma} \bra{n, \nu, \sigma'},
\end{equation}
where $m$ and $n$ index the unit cells along the axis of periodicity, $M$ is the number of unit cells, $\mu$ and $\nu$ are the sublattice indices, and $\mathcal{N}$ denotes the number of sublattices. Note that in this notation, $G^{\sigma\sigma'}_{m\mu,m\mu} = 0$. We now make the discrete Fourier transform $\ket{m, \mu, \sigma} = (1/\sqrt{M}) \sum_\k \exp{(i\k\cdot\mathbf{r}_{m\mu})} \ket{\k, \mu, \sigma}$ to arrive at
\begin{equation}
    H_{\mathrm{eff}} = \frac{\omega_0}{M} \sum_{m=1}^{M} \sum_{\mu=1}^\mathcal{N} \sum_{\sigma} \sum_{\k,\k'} e^{i(\k' - \k)\cdot\mathbf{r}_{m\mu}} \ket{\k, \mu, \sigma} \bra{\k', \mu, \sigma} - \frac{3}{2} \frac{\lambda_0 \Gamma_0}{M} \sum_{m,n=1}^{M} \sum_{\mu,\nu=1}^\mathcal{N} \sum_{\sigma,\sigma'} \sum_{\k,\k'} e^{-i\k\cdot\mathbf{r}_{m\mu}} e^{i\k'\cdot\mathbf{r}_{n\nu}} G_{m\mu,n\nu}^{\sigma \sigma'} \ket{\k, \mu, \sigma} \bra{\k', \nu, \sigma'}
\end{equation}
for $\k = k \khat$ and $k = \pi j/(Ma)$. Here, $a = \abs{\mathbf{r}_{m+1,\mu} - \mathbf{r}_{m\mu}}$ is the lattice spacing between adjacent unit cells and $j = 0,...,2M$ is an integer. Noting that the Green's tensor depends only on the relative coordinate $\mathbf{R}^l_{\mu\nu} \equiv \mathbf{r}_{m\nu} - \mathbf{r}_{n\nu}$ for $l = m-n$, the second term can be written as
\begin{equation}
    - \frac{3}{2} \frac{\lambda_0 \Gamma_0}{M} \sum_{l=-\infty}^{\infty} \sum_{m=1}^{M} \sum_{\mu,\nu=1}^\mathcal{N} \sum_{\sigma,\sigma'} \sum_{\k,\k'} e^{i(\k' - \k)\cdot\mathbf{r}_{m\mu}} e^{-i\k'\cdot\mathbf{R}^l_{\mu\nu}} G_{m\mu,l\nu}^{\sigma \sigma'} \ket{\k, \mu, \sigma} \bra{\k', \nu, \sigma'}.
\end{equation}
In the limit of large $M$, the identity
\begin{equation}
    \sum^M_{m=1} e^{i(\k'-\k)\cdot \mathbf{r}_{m\mu}} = M \delta_{\k\k'}
\end{equation}
yields the partially diagonalized Hamiltonian
\begin{equation}
    H_{\mathrm{eff}} = \sum_{\mu=1}^\mathcal{N} \sum_{\k} \sum_{\sigma} \omega_0 \ket{\k, \mu, \sigma} \bra{\k, \mu, \sigma} - \frac{3}{2} \lambda_0 \Gamma_0 \sum_{\k} \sum_{\mu,\nu=1}^\mathcal{N} \sum_{\sigma,\sigma'} \sum_{l=-\infty}^{\infty}  e^{-i\k\cdot\mathbf{R}^l_{\mu\nu}} G_{m\mu,l\nu}^{\sigma \sigma'} \ket{\k, \mu, \sigma} \bra{\k, \nu, \sigma'}.
\end{equation}
Changing notation slightly, we may drop the superscript on $\mathbf{R}^l_{\mu\nu}$ by replacing the sum over $l$ with a sum over the entire set of $\mathbf{R}_\mu = \mathbf{R}_{\mu\nu} - \mathbf{n}_{\mu\nu}$. Here, $\mathbf{R}_\mu$ denotes a Bravais lattice vector on sublattice $\mu$ and $\mathbf{n}_{\mu\nu}$ denotes the basis vector pointing from sublattice $\nu$ to sublattice $\mu$. Writing $G_{m\mu,l\nu}^{\sigma \sigma'}$ as $G_{\sigma\sigma'}(\mathbf{R}_\mu + \mathbf{n}_{\mu\nu})$, the Hamiltonian takes the form $H_{\mathrm{eff}} = \sum_\k H(\k)$, where the Bloch Hamiltonian
\begin{equation}
    H(\k) = \sum_{\mu,\nu=1}^{\mathcal{N}} \sum_{\sigma,\sigma'} h_{\mu\sigma,\nu\sigma'}(\k) \ket{\k, \mu, \sigma} \bra{\k, \nu, \sigma'}
\end{equation}
has matrix elements $h_{\mu\sigma,\nu\sigma'}(\k) = \omega_0 \delta_{\mu\nu} \delta_{\sigma\sigma'} +  \chi^I_{\mu\sigma,\nu\sigma'} + \chi^{II}_{\mu\sigma,\nu\sigma'}$ for
\begin{align}
    \chi^I_{\mu\sigma,\nu\sigma'} &= -\frac{3}{2}\lambda_0 \Gamma_0 \sum_{\mathbf{R}_{\mu} \neq 0} e^{-i\k\cdot\mathbf{R}_{\mu}} G_{\sigma\sigma'}(\mathbf{R}_{\mu})\delta_{\mu\nu}\\
    \chi^{II}_{\mu\sigma,\nu\sigma'} &=  -\frac{3}{2}\lambda_0 \Gamma_0 \sum_{\mathbf{R}_{\mu}} e^{-i\k\cdot (\mathbf{R}_\mu + \mathbf{n}_{\mu\nu})} G_{\sigma\sigma'}(\mathbf{R}_\mu + \mathbf{n}_{\mu\nu})(1 - \delta_{\mu\nu}).
\end{align}
Finally, to enforce periodicity of the Brillouin zone, it is necessary to apply the local gauge transformation \cite{bena_remarks_2009} $b_{\k\mu\sigma} \to e^{-i\k \cdot \mathbf{n}_{\mu}} b_{\k\mu\sigma}$. A redefinition of $H(\k)$ to include this $\k$-dependent phase transforms 
\begin{equation}\label{eq:gauged}
    \chi^{II}_{\mu\sigma,\nu\sigma'} \to -\frac{3}{2}\lambda_0 \Gamma_0 \sum_{\mathbf{R}_{\mu}} e^{-i\k\cdot\mathbf{R}_{\mu}} G_{\sigma\sigma'}(\mathbf{R}_{\mu} + \mathbf{n}_{\mu\nu})(1 - \delta_{\mu\nu})
\end{equation}
and ensures that $H(\k + \mathbf{G}_\mu) = H(\k)$ for any reciprocal lattice vector, $\mathbf{G}_\mu \cdot \mathbf{R}_\mu = 2\pi$.

\section{Spin dynamics under rotations and reflections\label{app:ortho}}

Because the elements of $O(3)$ preserve spatial distances and the spin-preserving interaction $G_{ij}^{\sigma \sigma}$ depends only on $r_{ij}$, it is sufficient to consider the effects of group multiplication on the spin-flip interaction alone. It is convenient to work in cylindrical coordinates and in the $(\dhat_1, \dhat_2, \qhat)$ basis such that $\mathbf{r}_i = \varrho_i \cos(\phi_i) \dhat_1 + \varrho_i \sin(\phi_i) \dhat_2 + r_{\q} \qhat$, where $\varrho_i$ is the radial coordinate of emitter $i$ and $\phi_i$ is the corresponding azimuthal coordinate measured in the polarization plane. Substituting the circular polarization vectors of Eq.~\eqref{eq:pols} into Eq.~\eqref{eq:J}, the spin-flip amplitude is then
\begin{equation}\label{eq:spin_flip}
        G_{ij}^{\uparrow \downarrow} = \frac{3 \lambda_0 \Gamma_0}{16\pi k_0^2 r_{ij}^5} e^{-2i(\phi_i + \phi_j)} \left (e^{i\phi_j} \varrho_i - e^{i\phi_i} \varrho_j \right )^2 e^{ik_0 r_{ij}}\left (-3 + k_0^2 r_{ij}^2 + i3k_0 r_{ij} \right ),
\end{equation}
with
\begin{equation}\label{eq:phi}
\phi_i = \tan^{-1}(\mathbf{r}_{i} \cdot \hat{\mathbf{d}}_2 / \mathbf{r}_{i} \cdot \hat{\mathbf{d}}_1).
\end{equation}
The inverse process $G_{ij}^{\downarrow \uparrow}$ is given by taking $\phi_i \to - \phi_i$ and $\phi_j \to - \phi_j$. Eq.~\eqref{eq:spin_flip} demonstrates that for $\mathbf{r}_{ij} \parallel \qhat$, the spin-flip interaction vanishes ($\varrho_i = \varrho_j$ and $\phi_i = \phi_j$). In this case, Eq.~\eqref{eq:general_ham} is diagonal in spin space and the dynamics are those of two uncoupled bosonic subspaces. If, on the other hand, the emitters are not collinear with the quantization axis ($\mathbf{r}_{ij} \nparallel \qhat$), then the spin-flip amplitude can be nonzero and lead to spin mixing. Note, however, that this is not a sufficient condition for nontrivial spin textures, which require the breaking of spin inversion symmetry (Section~\ref{sect:SOC}). 

The sign of \eqref{eq:phi} is the only quantity in the Hamiltonian that distinguishes $\uparrow$ from $\downarrow$. Thus, any spin dependent dynamics must be encoded in this phase, and orthogonal transformations that change the sign of this phase must map spin $\uparrow$ dynamics to spin $\downarrow$ dynamics and vice versa. We first consider proper rotations $R$, which satisfy $\det(R) = 1$.
\begin{definition}
    Let $R_b (\alpha)$ denote the orthogonal operator specifying azimuthal rotation by an angle $\alpha$ about an axis $b$.
\end{definition}
In the $(\dhat_1, \dhat_2, \qhat)$ basis, a rotation about the quantization axis has matrix representation
\begin{equation}
        \hat{R}_{\q} (\alpha) = \rcvect{\cos(\alpha) & -\sin(\alpha) & 0 \\
    \sin(\alpha) & \cos(\alpha) & 0 \\
    0 & 0 & 1}.
\end{equation}
Acting with this operator on an arbitrary lattice geometry yields the transformed position vectors
\begin{equation}
    \hat{R}_{\q} (\alpha) \mathbf{r}_i = \rcvect{\cos(\alpha) \mathbf{r}_i \cdot \dhat_1 - \sin(\alpha) \mathbf{r}_i \cdot \dhat_2 \\
    \sin(\alpha) \mathbf{r}_i \cdot \dhat_1 + \cos(\alpha) \mathbf{r}_i \cdot \dhat_2 \\
    \mathbf{r}_i \cdot \qhat} \equiv \mathbf{r}'_i.
\end{equation}
Invoking the identity $\tan^{-1}(u) \pm \tan^{-1}(v) = \tan^{-1}((u\pm v)/(1 \mp uv))$, the phase~\eqref{eq:phi} transforms as
\begin{align}\label{eq:rotate_q}
    \phi'_i &= \tan^{-1} \left ( \frac{\sin(\alpha) \mathbf{r}_i \cdot \dhat_1 + \cos(\alpha) \mathbf{r}_i \cdot \dhat_2}{\cos(\alpha) \mathbf{r}_i \cdot \dhat_1 - \sin(\alpha) \mathbf{r}_i \cdot \dhat_2} \right ) \nonumber \\
    &= \tan^{-1} \left ( \frac{\tan(\alpha) + \frac{\mathbf{r}_i \cdot \dhat_2}{\mathbf{r}_i \cdot \dhat_1}}{1 - \frac{\mathbf{r}_i \cdot \dhat_2}{\mathbf{r}_i \cdot \dhat_1} \tan{\alpha}} \right ) = \phi_i + \alpha.
\end{align}
Hence, rotations about $\q$ contribute only an overall phase to the spin-flip interaction and can be gauged away by a suitable redefinition of the $\phi_i = 0$ reference value.

We now consider improper rotations (reflections) $\R$, satisfying $\det(\R) = -1$. 
\begin{definition}
    Let $\R_\q (\alpha)$ denote the orthogonal operator specifying reflection through a plane containing $\q$ that makes an angle $\alpha$ with the $\mathbf{d}_1$ axis.
\end{definition}
The corresponding matrix representation is
\begin{equation}
    \hat{\R}_\q (\alpha) = \rcvect{\cos(2\alpha) & \sin(2\alpha) & 0 \\
    \sin(2\alpha) & -\cos(2\alpha) & 0 \\
    0 & 0 & 1}.
\end{equation}
Because rotations about $\q$ leave $\phi_i$ invariant (up to an arbitrary constant), it follows that all reflections $\R_\q (\alpha)$ have an equivalent effect on the spin dynamics. In more detail, we may write the combined rotoreflection operation as
\begin{equation}
R_{\q} (\alpha) \R_\q (\alpha') = \R_\q (\alpha' + \alpha/2) \equiv \R_\q (\alpha'').
\end{equation}
Then for $\mathbf{r}''_i \equiv \R_\q (\alpha'') \mathbf{r}_i$, the corresponding azimuthal coordinate in the polarization plane is given by
\begin{align}\label{eq:reflect}
    \phi''_i &= \tan^{-1} \left ( \frac{\sin(2\alpha'') \mathbf{r}_i \cdot \dhat_1 - \cos(2\alpha'') \mathbf{r}_i \cdot \dhat_2}{\cos(2\alpha'') \mathbf{r}_i \cdot \dhat_1 + \sin(2\alpha'') \mathbf{r}_i \cdot \dhat_2} \right ) \nonumber \\
    &= \tan^{-1} \left ( \frac{-\mathbf{r}_i \cdot \dhat_2}{\mathbf{r}_i \cdot \dhat_1} \right ) = -\phi_i,
\end{align}
where the second step follows by setting $\alpha' = -\alpha/2$. It follows that any reflection $\R_\q (\alpha')$ reverses the spin dynamics. 

In addition to reflections through planes containing the quantization axis, the sign of \eqref{eq:phi} can also be reversed under rotations about orthogonal axes. For simplicity, we consider here only rotations about the $\dhat_1$ axis, though those about other axes in the polarization plane follow similarly. In this case, the transformed vector $\mathbf{r}'''_i \equiv R_{\mathbf{d}_1} (\beta) \mathbf{r}_i$ has azimuthal coordinate 
\begin{equation}\label{eq:rotate_d}
    \phi'''_i = \tan^{-1} \left ( \frac{\cos(\beta) \mathbf{r}_i \cdot \dhat_2 - \sin(\beta) \mathbf{r}_i \cdot \qhat}{\mathbf{r}_i \cdot \dhat_1} \right ),
\end{equation}
and $\phi'''_i = -\phi_i$ for $\beta = \pi$.

\section{Topological classification\label{app:topo}}

For a non-Bravais lattice with $\mathcal{N}$ sublattices, the Hamiltonian \eqref{eq:k-space_H} gives rise to $2\mathcal{N}$ Bloch modes of the form $\ket{\psi_{n\k}} = e^{i\k\cdot\mathbf{r}} \ket{u_{n\k}}$, where $n$ is the band index. If the system admits a band gap, then the isolated $\mathcal{M}$ bands on one side of the gap obey the $U(\mathcal{M})$ gauge freedom 
\begin{equation}
    \ket{u_{n\k}} \to \sum_{m=1}^{\mathcal{M}} U_{mn}(\k) \ket{u_{m\k}},
\end{equation}
where the $\mathcal{M} \times \mathcal{M}$ unitary matrix $U_{mn}$ describes an equivalence class of physically identical Bloch manifolds. The non-Abelian Berry connection for each isolated manifold then follows as \cite{wilczek_appearance_1984, berry_quantal_1984}
\begin{equation}
    \mathbf{A}_{mn}(\k) = i \braket{u_{m\k} | \nabla_\k u_{n\k}}.
\end{equation}
For a closed loop $\mathcal{C}$ around the first Brillouin zone, the Berry phase is given by
\begin{equation}
    \varphi = \oint_{\mathcal{C}} \Tr{[\mathbf{A}(\k)]} \cdot d\k = -\Imag\left\{\ln\left[ \det \left(\mathcal{W}_{\mathcal{C}}\right)\right] \right\},
\end{equation}
where $\mathcal{W}_{\mathcal{C}}$ is the Wilson loop for the path $\mathcal{C}$ traversed in reciprocal space. In 1D and for discretized $k=k_0,...,k_L$, the Wilson loop may be written as \cite{gresch_z2pack_2017, vanderbilt_berry_2018}
\begin{align}\label{eq:wilson}
    \mathcal{W}_{\mathcal{C}} &= P\exp{ \bigg \{\oint_{\mathcal{C}} -i A_{mn}(k) dk \bigg \}} \nonumber \\
    &= \prod_{i=0}^{L-1} \exp{\{-i A^{(k_i,k_{i+1})}_{mn} dk\}},
\end{align}
where $P$ is the path-ordering operator and
\begin{align}
    \exp{\{-i A^{(k_i,k_{i+1})}_{mn} dk\}} &\approx \delta_{mn} - i A^{(k_i,k_{i+1})}_{mn} dk \nonumber \\
    &= \delta_{mn} + \bra{u_{mk_i}}(\ket{u_{nk_{i+1}}} - \ket{u_{nk_i}}) \nonumber \\
    &= M^{(k_i,k_{i+1})}_{mn}
\end{align}
for overlap matrix $M^{(k_i,k_{i+1})}_{mn} = \braket{u_{mk_i} | u_{nk_{i+1}}}$. Eq.~\eqref{eq:wilson} holds provided the cell-periodic functions are specified in the periodic gauge where $\ket{u_{nk_0}} = \ket{u_{nk_L}}$. In this case, the Wilson loop is easily computed as
\begin{equation}
    \mathcal{W}_{\mathcal{C}} = \prod_{i=0}^{L-1} M^{(k_i,k_{i+1})}_{mn}.
\end{equation}
Importantly, while $A_{mn}(k)$ is gauge-dependent, the 1D Berry phase (or Zak phase) is gauge invariant modulo $2\pi$. In general, the Zak phase for the 1D Bloch bands can assume any value, but is quantized to either $0$ (trivial) or $\pi$ (nontrivial) in the presence of either inversion or anti-inversion symmetry.

\bibliography{references}


\end{document}